\documentclass[twocolumn,aps,prd,amssymb,eqsecnum,nofootinbib,superscriptaddress,showpacs,floatfix]{revtex4}
\newlength{\picwidth}
\usepackage{epsfig}
\usepackage{amsmath}
\usepackage{dcolumn}
\usepackage{amssymb}
\usepackage{amsfonts}
\usepackage{enumerate}
\usepackage{bm}
\begin{document}
\def\be{\begin{equation}}
\def\ee{\end{equation}}
\def\ba{\begin{eqnarray}}
\def\ea{\end{eqnarray}}

\title{Luminosity distance in ``Swiss cheese" cosmology with randomized voids:\\ I. Single void size}

\author{R. Ali Vanderveld}
\affiliation{Jet Propulsion Laboratory, California Institute of Technology, Pasadena, CA 91109, USA}
\author{\'{E}anna \'{E}. Flanagan}
\affiliation{Center for Radiophysics and Space Research, Cornell University, Ithaca, NY 14853, USA}
\affiliation{Laboratory for Elementary Particle Physics, Cornell University, Ithaca, NY 14853, USA}
\author{Ira Wasserman}
\affiliation{Center for Radiophysics and Space Research, Cornell University, Ithaca, NY 14853, USA}
\affiliation{Laboratory for Elementary Particle Physics, Cornell University, Ithaca, NY 14853, USA}
\date{\today}
\begin{abstract}

Recently there have been suggestions that the Type Ia supernova data can
be explained using only general relativity and cold dark matter with
no dark energy.
In ``Swiss cheese'' models of the
Universe, the standard Friedmann-Robertson-Walker picture is
modified by the introduction of mass-compensating spherical
inhomogeneities, typically described by the
Lema\^{i}tre-Tolman-Bondi metric.  If these inhomogeneities correspond
to underdense cores surrounded by mass-compensating overdense shells,
then they can modify the luminosity distance-redshift relation in a way
that can mimic accelerated expansion.  It has been argued that this
effect could be large enough to explain the supernova data without
introducing dark energy or modified gravity.
We show that the large
apparent acceleration seen in some models
can be explained in terms of standard weak field gravitational
lensing together with insufficient randomization of void locations.
The underdense regions focus the light less than
the homogeneous background, thus dimming supernovae in a
way that can mimic the effects of acceleration.
With insufficient randomization of the spatial location of the voids
and of the lines of sight,
coherent defocusing can lead to anomalously large demagnification effects.
We show that a proper randomization of the voids and lines of sight
reduces the effect to the point that it can no longer explain the
supernova data.

\end{abstract}
\maketitle

\section{Introduction}

Supernova cosmology \cite{Riess, Perlmutter} has taught us that the
Universe is not well described by a homogeneous and flat matter
dominated model, with gravity governed by general relativity.  Type Ia
supernovae at a given redshift appear dimmer
than expected based on this simple model.  Many additions
to the model have been developed to explain this discrepancy,
including dark energy and modifications of general relativity on
cosmological length scales.  Another suggestion is that the
discrepancy results from large scale cosmological
inhomogeneities, and arises from an unjustified fitting of supernova
data to homogeneous models.
It is well known that inhomogeneity in
the cosmological matter distribution affects the redshifts and
apparent magnitudes of supernovae \cite{HG, CC, Lensing, V}, but could
such effects be large enough to trick us into thinking that the
Universe is accelerating when it is actually not?
This issue has been vigorously debated in the literature; see Ref.\
\cite{Celerier} for a review.

One way to address the inhomogeneity issue in a fully nonlinear and relativistic
fashion is with so-called ``Swiss cheese'' models \cite{HW, Sugiura, CMY1, LTB}.  In
these models, the homogeneous Friedmann-Robertson-Walker (FRW)
metric is modified by the introduction of
spherical regions of the spherically symmetric
Lema\^{i}tre-Tolman-Bondi (LTB) dust spacetime.
In this paper we will restrict attention to such models where
the FRW background is flat and matter dominated.
The spherical inhomogeneities must be mass
compensating and comoving in order to obtain a self consistent
solution with the LTB and FRW regions evolving independently.
The LTB regions can be of any size and can be located anywhere, so long as they
do not overlap.  However, for a realistic model of the Universe, the sphere locations should be randomized,
i.e.~they should not
placed on a lattice of any sort.  In this paper, we show that the
randomization issue is important and can have a dramatic
effect on the model's predictions.

Marra, Kolb, and Matarrese (hereafter MKM) \cite{MKM1, MKM2, Marra} use a Swiss cheese
model where the LTB spheres correspond to underdense voids surrounded
by overdense shells of matter.  They place these spheres on a regular
lattice, and compute numerically the luminosity distance-redshift
relation along a line passing through the spheres' centers.
They show that
the relation is modified enough to mimic a significant dark
energy density.

We show that the results of MKM can be explained to $\sim 10\%$ accuracy with the
formalism of standard weak field gravitational lensing.  The special
line of sight that they consider leads to a large dimming
because this is the line of sight with the lowest mean
density, corresponding to less focusing.  We also show that if the
void locations and lines of sight are properly randomized, then
the mean effect is very small,
in accordance with previous lensing
studies \cite{HW, Weinberg, Sereno2, KL}.
Using a Monte Carlo simulation, we compute the
statistical distribution of distance modulus shifts
at fixed redshift in the more realistic randomized scenario.

This paper is organized as follows.  Section II reviews
Swiss cheese models, focusing on the specific model constructed by
MKM.  Section III shows that the apparent acceleration found by MKM
is predominantly the result of gravitational lensing and of
insufficient randomization of the void locations.
In Sec. IV, we show that
randomizing the void locations reduces the effect significantly, to
the point that this model cannot explain the apparent acceleration
seen in the supernova data.  In Sec. V, we provide our concluding
remarks.

\section{Swiss cheese models}

The MKM model consists of a flat, matter dominated FRW background
with metric
\be
ds^2=-dt^2+a^2(t)\left(dr^2+r^2d\Omega^2\right)
\ee
and density $\rho_{\rm FRW}(t) \propto 1/a^3$.  From this FRW
background, spheres of constant comoving radius $R$ are removed and
replaced with spherically symmetric LTB dust solutions.  For simplicity we
describe this construction for a sphere centered at the origin, and for $r
\le R$.  The LTB metric in $r \le R$ is \cite{Bondi}
\be
ds^2=-dt^2+\frac{Y'^2(r,t)}{1+2E(r)}dr^2+Y^2(r,t)d\Omega^2~,
\ee
where $E(r)$ is the so-called energy function,
primes denote partial derivatives with respect to $r$ and
overdots denote partial derivatives with respect to $t$.
For the case $E(r)>0$ considered by MKM, the angular diameter distance
$Y(r,t)$ is given parametrically by
\be
Y(r,t)=\frac{M(r)}{12\pi E(r)}\left(\cosh u-1\right)~,
\label{Y1}
\ee
\be
t-t_b(r)=\frac{M(r)}{6\pi\left[2E(r)\right]^{3/2}}\left(\sinh u-u\right)~,
\label{Y2}
\ee
where the mass function $M(r)$ and the bang time function $t_b(r)$,
as well as the energy function $E(r)$, are freely specifiable functions.
Following MKM, we use units with $6 \pi G =1$.
The density of the dust is given by
\be
\rho(r,t) = \frac{M'(r)}{4 \pi Y'(r,t) Y(r,t)^2}~.
\label{density}
\ee

There are three conditions necessary for a consistent matching of the
LTB region with the surrounding flat FRW spacetime, satisfying the Israel
\cite{Israel} junction conditions \cite{Mansouri}.  The first is that the energy function
goes to zero at the boundary
\be
E(R)=0~.
\label{match1}
\ee
The second is that the bang time function goes to zero at the boundary
(assuming that the big bang occurs at $t=0$ in the FRW background)
\be
t_b(R)=0~.
\label{match2}
\ee
The third is that the LTB region be mass compensating, meaning that the enclosed mass
has to equal the mass originally excised from the FRW background.
This requirement is that
\be
M(R) = \frac{4 \pi}{3} \rho_{{\rm FRW},0} R^3
\label{match3}
\ee
where the constant $\rho_{{\rm FRW},0}$ is defined by $\rho_{\rm FRW}(t)
= \rho_{{\rm FRW},0} / a(t)^3$.  We also choose the scale factor $a(t)$ to equal $1$ today.

In addition to satisfying the Israel junction conditions, it is convenient to
specialize to models for which the coordinates $(t,r)$ match up in a
$C^1$ manner across the boundary $r=R$.  This guarantees that the
basis vectors $\partial/\partial t$ and $\partial / \partial r$ will
be continuous across $r=R$, and justifies our use of the same notation
$(t,r)$ for the coordinates outside and inside the sphere.
It can be shown that the necessary and sufficient conditions for this
continuity are
\be
M'(R) = 3 M(R)/R~,\ \ \ \ t_b'(R) = 0~, \ \ \ \ \ E'(R) = 0~.
\label{constraints1}
\ee

MKM specialize to a subclass of LTB models defined by imposing, at
some initial time $t=t_i$, the two conditions
\be
Y(r,t_i) = a(t_i) r~, \ \ \ \ {\dot Y}(r,t_i) = {\dot a}(t_i) r~.
\ee
These conditions fix two of the three functional degrees of freedom of
the LTB model, one of which is gauge.  MKM parameterize the remaining,
nongauge, functional degree of freedom in terms of the density
profile at the initial time $\rho_0(r) = \rho(r,t_i)$.
The mass inside radius $r$ is given in terms of this density profile by
\be
M(r)=4\pi a(t_i)^3 \int^r_0\rho_0({\bar r}) {\bar r}^2d{\bar r}~.
\label{mass}
\ee
It is convenient to parameterize this as
\be
M(r)=M_{\rm FRW}(r)\left[1+\delta(r)\right]~,
\label{delta0}
\ee
where
\be
M_{\rm FRW}(r)=\frac{4\pi}{3}r^3 a(t_i)^3 \rho_{\rm FRW}(t_i)
\label{MFRW}
\ee
is the mass enclosed that would apply in the FRW background, and
$\delta(r)$ is the fractional mass excess.
The energy function $E(r)$ and bang time function $t_b(r)$ can be
written in terms of the fractional mass excess as
\be
E(r) = - \left[ \frac{ M_{\rm FRW}(r) }{6 \pi a(t_i) r} \right]
\delta(r),
\label{energy}
\ee
and 
\be
t_b(r)=t_i\left\{1+\frac{3}{2\delta(r)}-\frac{3\left[1+\delta(r)\right]}
{2\delta(r)}\frac{\tanh^{-1}\left[\sqrt{-\delta(r)}\right]}{\sqrt{-\delta(r)}}\right\}~.
\label{bangtime}
\ee
The last formula is valid only for $\delta <0$, the case considered by MKM.
From the results (\ref{mass}), (\ref{delta0}), and (\ref{bangtime}), it
follows that all of the matching and continuity conditions (\ref{match1}) --
(\ref{constraints1}) will be satisfied if we choose the fractional mass excess
$\delta(r)$ to satisfy
\be
\delta(R) = 0~, \ \ \ \ \delta'(R) = 0~.
\ee
The first of these conditions is the mass compensation condition,
while the second is a requirement that the density be continuous
across $r=R$.

In the next section, we will
we make the approximation of using the relativistic LTB density
profile $\rho(r,t)$ as an input to the standard weak field lensing
formalism.  To summarize the discussion above, the procedure for
computing the density $\rho(r,t)$ from a specified initial density
profile $\rho_0(r)$ is as follows.  (i) Compute the mass enclosed $M(r)$ using
Eq.\ (\ref{mass}) and the fractional mass excess $\delta(r)$ using Eqs.\ (\ref{delta0})
and (\ref{MFRW}).  (ii) Use the fractional mass excess to compute the
energy function (\ref{energy}) and bang time function (\ref{bangtime}).
(iii) Then compute the angular diameter distance $Y(r,t)$ from Eqs.\
(\ref{Y1}) -- (\ref{Y2}), and finally insert this into the expression
(\ref{density}) for the density $\rho(r,t)$.

\begin{figure}
\begin{center}
\epsfig{file=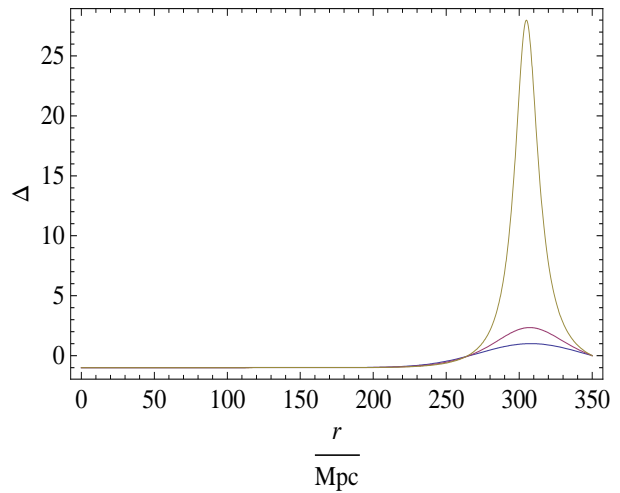,width=\picwidth,height =6.5cm}
\caption{The density contrast $\Delta(r,t)=[\rho(r,t)-\rho_{\rm
    FRW}(t)]/\rho_{\rm FRW}(t)$ as a function of LTB coordinate
  $r$ in the model of MKM at the times $t=0.2$
  (blue, bottom curve), $0.6$ (pink, middle curve), and $1$ (yellow, top curve).}
\label{delta}
\end{center}
\end{figure}

For the initial density profile for the LTB patches, MKM choose the Gaussian profile
\be
\rho_0(r)=A\rho_{\rm FRW}(t_i)\left\{\exp\left[-B\left(r-r_0\right)^2\right]+C\right\}
\ee
with constants $A\approx 2$, $B\approx 4\times 10^{-4}~{\rm
  Mpc}^{-2}$, $r_0\approx 308~{\rm Mpc}$, and $C\approx 5\times 10^{-5}$. This
is the density at the initial time $t_i=0.2$, where units are chosen
such that the big bang occurred at $t=0$ and today is at $t=1$, implying that the scale factor is $a(t)=t^{2/3}$.  
Here, $r$ is in Mpc and
the present-day radius of an LTB patch is $R=350~{\rm Mpc}$.  We will use an
initial density profile that
is very similar to that of MKM for all of
our analyses below, except that we slightly
modify these constants to accurately enforce mass conservation, while setting $C=0$ for simplicity.  Such small changes affect our results only negligibly.
We plot the resulting density contrast
$\Delta(r,t)=[\rho(r,t)-\rho_{\rm FRW}(t)]/\rho_{\rm FRW}(t)$ in
Fig.~\ref{delta} for $t=0.2$, $0.6$, and $1$.

\section{The results of Marra, Kolb, and Matarrese}

In the MKM model, the void regions are placed on a regular cubic
lattice wherein the voids are just touching.  The chosen line of
sight is through the center of a sequence of voids;
see Fig.~2 of Ref.~\cite{MKM2}
for a sketch of their setup.  The void size is such that the distance
through 5 voids corresponds to $z\approx 1.8$.  After numerically
integrating the fully relativistic equations for light rays, MKM
display results for the change in the redshift and
distance modulus, both as a function of source redshift, in Figs.~4
and 5 of \cite{MKM2}.  They find that there is a negligible effect on
the redshift and there is a significant effect on the distance
modulus, with supernovae systematically becoming demagnified.  This
demagnification increases with the number of void crossings.
The effect is large enough that supernovae that are 5 voids away
suffer a distance modulus shift of about $0.34$ mag, and this suffices
to move them to the Hubble diagram of a $\Omega_M=0.6$,
$\Omega_{\Lambda}=0.4$ universe.

We now argue that this result is predominantly due to
gravitational lensing.  Although the inhomogeneities are mass
compensating, light beams passing through the spheres' centers
encounter less mass than beams that intersect the spheres with larger
impact parameters.  Therefore, centered beams are focused less than
they would be in the FRW background.

In weak lensing theory, the net magnification $\mu$ produced by a density
perturbation $\Delta \rho({\bf x},\eta)$ is $\mu = (1 - \kappa)^{-2}$.
Note that we are neglecting shear, which gives a contribution to $\mu$ of the same order as the $\kappa^2$ corrections.
The lensing convergence $\kappa$ is given by the integral along
the line of sight \cite{Durrer}
\be
\kappa = 4 \pi G \int_{\eta_S}^{\eta_O} d \eta \, \frac{ (\eta - \eta_S)
  (\eta_O - \eta)}{\eta_O - \eta_S} a(\eta)^2 \Delta \rho[{\bf x}(\eta),\eta]~.
\ee
Here the integral is with respect to conformal time $\eta$
and is along the unperturbed ray ${\bf x}(\eta)$, and
$\eta_S$ and $\eta_O$ are the values of conformal time at the source
and at the observer.  Rewriting this integral in terms of the comoving
distance $x = \eta_O -\eta$, expressing it in terms of the density
contrast $\Delta({\bf x},\eta) = \Delta \rho / \rho_{\rm FRW}$,
and eliminating $\rho_{\rm FRW}$ using the Friedmann equation gives
\be
\kappa = \frac{3}{2} H_0^2 \int_0^{x_S} dx \, \frac{ x(x_S -
  x)}{x_S a(\eta_0 - x)} \, \Delta( {\bf x}(x), \eta_0 - x)~,
\ee
where $x_S = \eta_S - \eta_O$.  Next, since the individual spheres are small compared
with the horizon scale, we can treat the first factor in the integrand
as constant across each sphere, and we can also neglect the time
dependence of $\Delta$ within each sphere to leading order.  This
gives
\be
\kappa = \frac{3}{2} H_0^2 \sum_j \, \frac{ x_j(x_S - x_j)}{x_S
  a(\eta_j)}\, {\cal I}_j~,
\label{kappaf}
\ee
with
\be
{\cal I}_j = \int_{{\cal S}_j} dx \,
\, \Delta( {\bf x}(x), \eta_j)~,
\label{kappaf1}
\ee
where $x_j$ is the
value of $x$ at the
point of closest approach of the ray to the center of sphere $j$, and
$\eta_j = \eta_0 - x_j$ is the conformal time at that point of closest approach.
The interval ${\cal S}_j$ is the
intersection of the ray with the $j$th sphere, given by $|x - x_j|
\le \sqrt{R^2 - b_j^2}$, where $b_j$ is an impact parameter.
Rewriting the integral (\ref{kappaf1}) in terms of the comoving radial
coordinate ${\bar r} = \sqrt{b_j^2 + (x-x_j)^2}$ gives
\be
{\cal I}_j = 2 \int_{b_j}^R d{\bar r} \, \frac{{\bar r}}{\sqrt{{\bar
      r}^2 - b_j^2}} \, \Delta({\bar r},\eta_j)~.
\ee
To evaluate this integral, we use the relation ${\bar r} =
Y(r,t_j)/a(t_j)$ between the comoving radial coordinate ${\bar r}$ and
the LTB radial coordinate $r$, where $t_j = t(\eta_j)$.
Finally the distance modulus shift $\Delta m$ is related to the lensing
convergence $\kappa$ by
\be
\Delta m=5\log_{10}\left(1-\kappa\right)~.
\label{deltam}
\ee

We have computed the distance modulus shifts $\Delta m$ using
Eqs.\ (\ref{kappaf}) and (\ref{deltam}) for the model of MKM, with our aforementioned slight modifications. We also use $H_0=70$ km/s/Mpc. These results are summarized in Table~\ref{dms}.  Comparing with
Fig.~5 of Ref.\ \cite{MKM2}, we see that our weak field results reproduce
their fully relativistic results to $\sim 10\%$ accuracy.  This is consistent with what one would
expect given the approximations that we have made, since $R H_0 \sim 0.1$ for $R = 350$ Mpc and $H_0=70$ km/s/Mpc.  We conclude that gravitational lensing produced by the coherent
arrangement of voids provides a simple physical explanation for the
phenomenon seen in their analysis.
\begin{table}[h]
\caption{Supernova distance, in terms of the number of voids in
  between source and observer, and our result for the distance modulus
  shift $\Delta m$ if the voids are in a lattice.  The void radius is
  $R = 350$ Mpc.}
\begin{ruledtabular}
\begin{tabular}{cc}
Distance (in voids) & $\Delta m$ \\
\hline
1 & 0.015 \\
2 & 0.047 \\
3 & 0.11  \\
4 & 0.20  \\
5 & 0.33
\end{tabular}
\end{ruledtabular}
\label{dms}
\end{table}

We can get a good estimate the coherent MKM effect by using the Dyer and Roeder \cite{DR, Sereno1} formalism.  This is based the work of Zel'dovich \cite{Zeldovich}, wherein he considered cosmological observations in a flat matter dominated universe where there was no matter inside any observed light beams, and hence no gravitational focusing.  Ignoring the effects of shear, the observed flux from a source at redshift $z$
in such a model is $F_{empty}$, where
\be
{F_{empty}\over F_{filled}}={25[1-(1+z)^{-1/2}]^2\over
(1+z)^2[1-(1+z)^{-5/2}]^2}~,
\ee
with $F_{filled}$ being the flux for a filled beam. Thus, high redshift sources would be dimmed.
The corresponding magnitude shift is
\be
\Delta m_{empty}=5\log_{10}\left\{{(1+z)[1-(1+z)^{-5/2}]\over
5[1-(1+z)^{-1/2}]}\right\}~,
\ee
and for a model in which we observe through a complex of voids,
this must be an upper bound to the magnitude shifts possible.
For $z=1.8$, we find $\Delta m_{empty}=0.546$.  In the absence of shear, no arrangement of matter along the line of sight can produce a magnitude shift larger than this.

Dyer and Roeder generalized this analysis to 
partially filled beams. They found
\be 
\Delta m_\alpha=5\log_{10}\left\{{(1+z)^{(\beta-1)/4}
[1-(1+z)^{-\beta/2}]\over\beta[1-(1+z)^{-1/2}]}\right\}~,
\ee
where $\beta=\sqrt{25-24\alpha}$ and $\alpha$ is the filling
factor of the beam: $\alpha=1$ corresponds to a flat dark matter
dominated model and $\alpha=0$ corresponds to empty beams.

The limiting case is $\alpha=0$ for empty beams, but what is the
limiting case for passage through a mass compensated void? 
Consider an extreme example in which the void has all of its mass
concentrated in a shell at its boundary. Then if the mass is $M$ and the radius is
$R$, the integrated column density through the sphere is
$M/2\pi R^2={2\over 3}\rho R$, i.e.~this is zero except for the shell,
which is pierced twice. This is $1/3$ of the value for a smooth
sphere, and hence for this model $\alpha=1/3$ would be appropriate.
Using this value we get $\beta=\sqrt{17}$ and 
\be
\Delta m_{1/3}=5\log_{10}\left\{{(1+z)^{(\sqrt{17}-1)/4}
[1-(1+z)^{-\sqrt{17}/2}]\over\sqrt{17}[1-(1+z)^{-1/2}]}\right\}~.
\ee
For $z=1.8$, this gives $\Delta m_{1/3}=0.369$.  This just slightly overestimates the MKM result of $\Delta m\approx 0.34$.

\section{Void randomization}

As we have argued, the coherent arrangement of the voids on a lattice
in the model of MKM artificially boosts the demagnification effect.  In
this section, we show that the net effect of the voids is small if the
void locations relative to the line of sight are sufficiently
randomized.

As is well known, the mean magnification due to gravitational lensing
should be negligible on average in models such as ours with small,
mass-compensating inhomogenities that are randomly distributed \cite{Weinberg, HW, KL}.  
We can
understand this result in a different way with a simple argument.
First, consider a spherically-symmetric matter distribution with a
total radius $R$, a total mass $M$, and a radius-dependent density
$\rho(r)$.  Let us further assume that a small beam comes in from
the $z$ direction and then encounters the sphere with an impact parameter
$b$, which will then be in the $x-y$ plane. The magnification of the
beam is proportional to the integrated column density along the beam's
unperturbed path, which is defined to be
\be
\Sigma \equiv \int_{path} \rho\left[r(\lambda)\right] d\lambda~,
\ee
where the path depends on $b$, and $\lambda$ is the affine parameter
of a central ray.  Since we are using the unperturbed path for this,
it is clear that we do the above integral along a straight line,
parallel to the $z$ axis.  Then the total mass of the spherical
distribution comes from integrating $\Sigma$ over the remaining two
spatial axes, the $x$ and the $y$:
\ba
M&=&\int_{all~space}\rho(r) dV = \int_{all~space}\rho(r) dx dy dz\nonumber\\
&=& \int_{x-y~plane}\Sigma\left[b(x,y)\right] dx dy \nonumber\\
&=& \int^R_0 \Sigma(r) 2\pi r dr~.
\ea
The average focusing resulting from a spherical mass distribution is
proportional to the average over impact parameters of the integrated
column density,
\ba
\langle \Sigma \rangle &=& \frac{1}{R^2}\int_0^R \Sigma(b)\times 2b db = \frac{1}{R^2}\int_0^R \Sigma(r) 2r dr\nonumber\\
&=&\frac{1}{R^2}\left(\frac{M}{\pi}\right) = \frac{M}{\pi R^2}
\ea
which only depends on total mass and size of the spherical region, and
not on how the mass is distributed.  Thus, we see that removing a
sphere of FRW spacetime and replacing it with a mass-compensating LTB
patch, as we did in the model discussed above, will not affect the
magnification on average.  This is true within the domain of validity
of weak lensing theory.
The distance modulus shift depends on the
logarithm of the magnification and thus it is not necessarily going to
have a vanishing mean, but we nonetheless find that it is small in the
model considered here.

\begin{figure}
\begin{center}
\epsfig{file=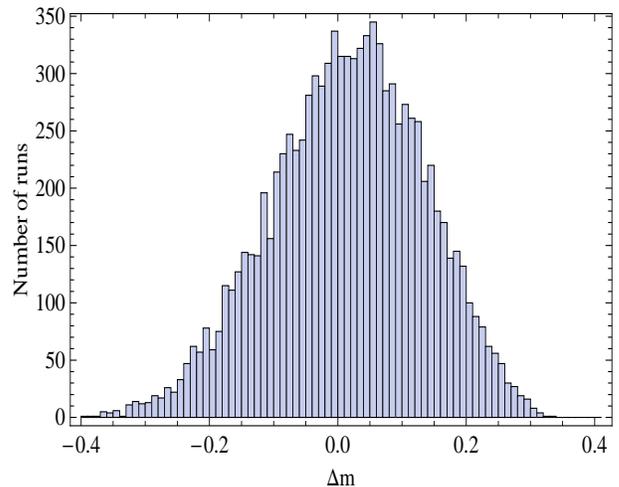,width=\picwidth,height =6.5cm}
\caption{A histogram of the distance modulus shifts $\Delta m$ found for $10,000$ realizations of our randomized void scenario}
\label{histogram}
\end{center}
\end{figure}

To find the statistical distribution of the distance modulus shifts
that one would expect after passage through 5 voids in a randomized
version of this Swiss cheese model, we do a Monte Carlo analysis.  In
this analysis, we pick random impact parameters for each of the $5$
void passages, with $b^2$ being uniformly distributed between $0$ and
$R^2$.  We find that the statistical distribution of distance modulus
shifts has a mean of $-0.003$ mag and a standard deviation of $0.11$
mag.  
Note that we have neglected shear, which might
change the mean magnitude shift by a factor of order $2$.
Figure~\ref{histogram} shows the distribution of these shifts
found for $10,000$ realizations of voids with random impact parameters
relative to the line of sight to the light source.

\section{Discussion}

MKM found significant changes to
the distance modulus of distant supernovae in a particular Swiss cheese
model.  We have argued that the large size of their effect stems from the fact
that their voids are on a regular lattice,
with the line connecting observer and source passing straight through
the void centers.
Their result can be reproduced with $\sim 10\%$ accuracy using weak
field gravitational lensing, and can be explained by noting that the
integrated column density is lowest
along lines of sight that pass through the centers of the voids.
The lower column density leads to less focusing and thus dimmer supernovae.

We also showed that in a more
realistic Swiss cheese model, the mean magnification due to
gravitational lensing is very small.
We simulated the distribution of distance modulus shifts obtained from
a model where the light rays no longer pass only through the centers
of the voids, but instead encounter them with random impact parameters.  For $N=5$ voids
out to $z=1.8$, MKM obtained a coherent distance modulus shift of $0.34$
mag, whereas our randomized model gives a mean of $-0.003$ mag
and a standard deviation of $0.11$ mag.  This mean is small and
this standard deviation is comparable to the intrinsic scatter of Type
Ia supernovae, and thus this particular model cannot explain the supernova data.  An analysis of the luminosity distance-redshift relation in a scenario with randomly-sized voids is the subject of future work.

While our results clarify the significance of the special setup underlying
the specific model of MKM, they do not exclude the possibility that
inhomogeneties may account for some of the apparent acceleration of the
Universe. However, our work does suggest that coherent underdense structures
on scales of order the horizon size
generally are needed to obtain large apparent accelerations.
Examples of models with such very large-scale structures are the spherical void models of
Refs.\ \cite{Tomita, Alnes, VFW, Garfinkle, CMY2}.
A general argument against such models is fine tuning: one would generically expect
large anisotropies in the luminosity distance at a given redshift in any model
with a large apparent acceleration, and the observational limits on anisotropy are small.
Going beyond fine tuning, there are limits on large spherical void models from current observations of the cosmic microwave background spectrum \cite{CMB} and low redshift Type Ia supernovae \cite{Ia},
and further constraints could come from future observations of baryon acoustic oscillations \cite{BAO} and time drifts of cosmological redshifts \cite{zdot1, zdot2}.

We also note that in spherical void models, fitting luminosity distance data to spatially flat cosmological models would
generally suggest the presence of dark energy.  If we allow more general fits to non flat models,
then we will get a fit with non zero curvature and non zero dark energy, with the relative proportion depending on the details of the model. In linear theory, the shrinking mode of an LTB model is the same as the mode parameterized by the bang time function,
and the growing mode is the mode parameterized by the energy function, which parameterizes
curvature.  Therefore in a model with no important shrinking mode component, it is likely that
the best fit will be have a higher proportion of curvature than dark energy.

\begin{acknowledgments}
The work of RAV was carried out at the Jet Propulsion Laboratory, California
Institute of Technology, under a contract with NASA. We also
acknowledge the support of NSF Grant Nos. PHY-0457200 and
PHY-0555216, and NASA Grant No. NNX08AH27G.
\end{acknowledgments}

\end{document}